# Segregation patterns in binary mixtures with same layer-thicknesses under vertical vibration

Yahya Sandali[1], Cheng Xu[1,*], Ning Zheng[1,2,**], Gang Sun[3,**], Qingfan Shi[1,2,**]

[1]School of Physics, Beijing Institute of Technology, Beijing 100081, China

[2]Key Laboratory of Cluster Science of Ministry of Education, Beijing 100081, China

[3]Key Laboratory of Soft Matter Physics and Beijing National Laboratory for Condensed Matter Physics,

Institute of Physics, Chinese Academy of Sciences, Beijing 100190

**E-mail: gsun@iphy.ac.cn, ningzheng@bit.edu.cn and qfshi123@bit.edu.cn

*This author should be considered as co-first authors.

## Abstract

Inspired by the theoretical prediction [Phys. Rev. Lett. 86, 3423 (2001)] and the disputed experimental results [Phys. Rev. Lett. 89, 189601(2002), Phys. Rev. Lett. 90, 014302 (2003)], we systematically investigate the pattern of binary mixtures consisting of same layer-thickness under vertical vibration. Various kinds of mixtures with different diameters and densities particles are used to observe the separation regime. It is found that these mixtures behave like five kinds of segregation patterns for different driving control parameters, i.e., Brazil nut (BN), reversed Brazil nut (RBN), Mixed states, light-BN (LBN), and light-RBN (LRBN), where the latter two regimes are neither purely segregated nor completely mixed states. Not only that, but LBN (LRBN) is observed to be the transition path from BN (RBN) to Mixed state. Moreover, BN phenomenon takes place in the area of low density ratio and found to be independent of layer structure, while RBN is sensitive to the layer structure and occurs at the large density ratio but lower diameter ratio. Our result may be helpful for the establishment of theory about the segregation and mixing of granular mixtures.

**Keywords**: segregation, granular material, reversed Brazil nut, layer thickness

## 1. Introduction

It is well known that, segregation of granular mixtures of different sizes is a common occurrence. When a particle mixture is shaken vertically or jostled, larger particles generally rise to form a layer on the top while the smaller particles filter down to form a layer on the bottom, known as the Brazil nut segregation (BN). It can just lead to substandard products in many industrial processes such as powder metallurgy, transportation, pharmaceutical production, food manufacturing, and concrete mixing, etc.[1-3], granular separation phenomenon has been an attractive area of research in the past eighty decades. In addition to engineering applications, the mechanism of separation is also an important concern to the researchers. The most commonly quoted two models are that, the ascent of larger particle is due to the falling of smaller particles into the voids produced underneath the larger particle after each shaking cycle [4-7]; global convection drives the larger ones to the top of mixture whereas small ones tend to sink to the bottom [8,9]. The other researches stated that, arching [10], inertia[11, 12], buoyancy [13], friction [14], pressure [15], density drive [16-19] are crucial elements in explaining BN. For further reveal the puzzle of mixture segregation, Hong et al. proposed a competition mechanism between the percolation and the condensation of particles, and consequently predicted the existence of the reversed BN (RBN) where the large particles sink to the bottom in the system consisting of same layer-thickness [20]. Noteworthy here, the system of same layer-thickness is one of fundamental assumption in Hong's theory. From then on, the occurrence condition of RBN began to face the challenges. Breu et al. [21] claimed that 82% of experimental data confirm this prediction by deliberately adjusting the amplitude and frequency of vertical vibration, while the other studies could not verify the prediction of Hong et al. experimentally [22, 23] and numerically [14]. For example, Canul-Chay et al did not find evidence of RBN although they carried out many experiments according to Hong's conditions such as intermediate temperature and same layer-thickness.

Obviously, the segregation pattern of binary mixtures is still a controversial problem. In fact, the condensation mechanism can only be observed for an intermediate granular temperature, which is difficult to attain experimentally due to its gradient existence along vertical vibration container [22]. From the point of view of industrial application, the system is totally controlled by the vibration frequency and amplitude instead of strictly considering the temperatures of species, i.e., the system is controlled arbitrarily. Furthermore, almost all of the experiments for segregation in binary mixtures take the driving parameters arbitrarily without considering the temperatures of both species until now. What's more, the driving parameters are arbitrarily exerted on the samples has more practical significance. Based on this analysis, it is interesting to question what the same-layer system would behave under the arbitrarily excitation. To our knowledge so far, this open problem has not yet been well explored. Therefore, further investigation is needed if we want to uncover the physics of segregation phenomena. In addition, better understanding of segregation conditions helps make it possible to tune, avoid or invert the segregation patterns of granules.

In this paper, we focus on the investigation of binary mixtures with same layer-thickness under vertical vibration. By controlling the driving parameters such as dimensionless acceleration and frequency, and using wide range of densities, we try to experimentally observe the behaviors of system and verify whether RBN phenomenon occurs without strictly take the temperature condition as Hong et al suggested. Furthermore, we study how the separation patterns vary depending on their diameter and density ratios of two species, and also the transition path from segregation pattern to mixed state. We expect that the present results could be helpful for establishing new segregation model of binary mixtures.

## 2. Experimental setup

A vertical vibration system was used to investigate the segregation behaviors of the binary mixtures. The container is a cylindrical glass cell of inner diameter 36 mm and height150 mm, which is vertically placed and fixed on a machined flat copper substrate. This flat substrate is supported on the horizontal surface of an electromechanical shaker (JZK-60T) which moves vertically with the ratio of the horizontal vibration amplitude to the vertical vibration amplitude less than 5%. The vertical harmonic displacement function is $A\sin(2\pi ft)$, where $A$ and $f$ are the amplitude and frequency of the vibration, respectively. In general, vibration frequency $f$ and dimensionless acceleration amplitude $\Gamma = 4\pi^2 Af^2/g$ were used as control parameters, where $g$ is the acceleration due to gravity. Our experimental apparatus can work in the range of acceleration $\Gamma$ from 1.0 to 8.0 and range of frequency $f$ from 10 to 100 Hz. Some previous works have shown that the air in the container may play complicated roles in the segregation process [15, 24]. To eliminate the influence of air effectively, the container was sealed and evacuated by a mechanical pump to an air pressure below 50 Pa [25]. In addition, in order to reduce the accumulation of static charge, beads were well stirred before they were poured into the glass cylinder.

We prepared many granules with different densities and sizes of granules. The particle densities are ranging from 1.31 to 18.0 $g/cm^3$, and the particle sizes used in the experiment range from 0.25 to 3.0 mm (see Table I).

TABLE I. Granular particles used in experiments

| Material | Density (g/cm$^3$) | Diameter (mm) |
|---|---|---|
| Aluminum oxide | 1.3 | 0.25-3.0 |
| Magnesium | 1.7 | 0.32-0.7 |
| Glass | 2.5 | 0.25-3.0 |
| Zirconium oxide | 2.9 | 0.25-1.13 |
| Zirconium silicate | 3.8 | 0.25-1.13 |
| Titanium alloy | 4.5 | 0.32-0.7 |
| Ferrous alloy | 7.4 | 0.25-0.7 |
| Cobalt-chromium-molybdenum alloy | 8.4 | 0.25-1.13 |
| copper | 8.7 | 0.25-0.43 |
| Tungsten alloy 1 | 13.8 | 0.55 |
| Tungsten alloy 2 | 18.0 | 0.32-0.7 |

The segregation process may also be influenced by the amount of granules or the number of layers of the granules piled in the container. In this work, the system used consists of same layer-thickness $\mu$ of two species, i.e $\mu_L = \mu_S$ (the subscripts L and

S refer to large and small particles). We initially arranged the two granular species to an appropriate layer thickness (in unit d), so as to avoid the BN or RBN phenomenon being completely destroyed. The experiment started by the first stirring of the particles in the binary mixtures randomly. The container of the mixture was then vertically vibrated at fixed values of $\Gamma$ and $f$. The value $\Gamma$ of each experimental run changed by steps of $\Delta\Gamma = 0.1$. In each run we monitored the pattern for at least half an hour and record the $\Gamma$ and $f$ of each steady pattern.

### 3. Result and Discussion

For the system of same layer-thickness, we first review the theoretical research related so as to better understand our experimental investigation. Hong et al. [20] first predicted the existence of the RBN, and proposed a crossover condition from BN to RBN based on a competition mechanism between the percolation and the condensation of particles as the following

$$\frac{\rho_L}{\rho_S} \approx \frac{d_S}{d_L}, \qquad (1)$$

where $d_L, d_S, \rho_L, \rho_S$ are the diameter and the density of large and small particles, respectively. If the density ratio is smaller than the inverse of the diameter ratio the system should behave RBN and vice versa. Moreover, the occurrence of RBN has a precondition such as the system temperature $T$ is in a intermediate state $T_C^S < T < T_C^L$, where $T_C^S$, $T_C^L$ are the critical temperatures of small and large particles species, respectively. The critical temperature means that the state of system is fully fluidized. However, Canul-Chay et al. [22] could not observe RBN phenomenon, though they claimed that they run experiments according to Hong assumptions with system of *same layer-thickness* and at different temperatures between $T_C^S$ and $T_C^L$. Another experimental work argued that they could verify equation (1) and observe RBN although the layer thickness and the temperatures of two species were not measured directly and therefore no definite conclusion can be drawn from the published data [21,26]. It brings to the surface why Canul-Chayet al. could not observe while Breu et al. did so. Thus, it is an interesting question if we run an experiment for the binary mixture consisting of same layer-thickness by using more realistic exciting method instead of strictly considering the intermediate temperature.

We carried out a sequence of experiments by adjusting the vibration acceleration from zero to maximum value, meanwhile frequency value was fixed. The same procedure was repeated at fixing acceleration with changing frequency in the space of acceleration-frequency parameters. Each individual observation started with the generation of a mixed state by shaking 10 s at f =18 Hz and $\Gamma$=5, and then the sample was shaken with a specific $f$ and $\Gamma$ until a steady state was reached. Five kinds of segregation patterns were observed. In addition to BN, Mixed and RBN states as shown respectively in Fig. 1 (a) (c) and (e), two partially segregated states that we call them the light-BN (LBN) and light-RBN (LRBN) were observed (see Fig. 1(b) and (d) ). They are neither purely segregated nor completely mixed states. It was observed that these two states have the characteristic that lighter particles tend to rise and form a pure layer on top of the system while the heavier particles and some lighter ones stay at the bottom and form a mixed layer [17].

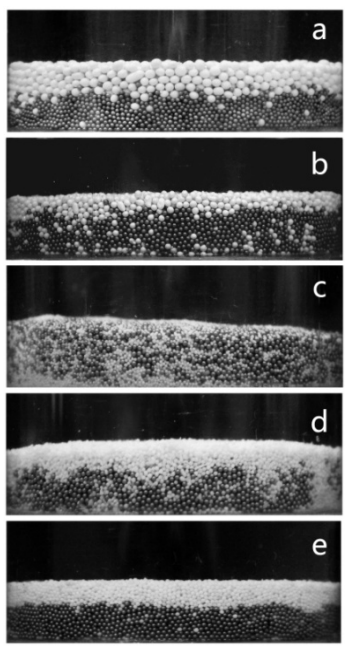

Fig.1 Five kinds of segregated patterns: (a) BN, 1.12 mm Zirconium silicate and 0.55 mm Cobalt-chromium-molybdenum alloy; (b) LBN, 0.76 mm Zirconium silicate and 0.55 mm Cobalt-chromium-molybdenum alloy; (c) Mixed, 0.55 mm Cobalt-chromium-molybdenum alloy and 0.32 mm Aluminum oxide；(d) LRBN, 0.55 mm Cobalt-chromium-molybdenum alloy and 0.425 mm Aluminum oxide; (e) RBN, 0.55 mm Cobalt-chromium-molybdenum alloy and 0.475 mm Aluminum oxide.

To give a comprehensive understanding the influence of controlling parameters on the segregation patterns, we plot a typical schematic phase diagram in $f - \Gamma$ parameter space in Fig.2, where the mixture used is the combination of 0.24 mm copper and 0.16 mm glass beads. The solid lines are drawn for eye guidance of the boundaries between the different segregation regimes. At low amplitudes ($\Gamma < 1.2$) the particles exhibit heaping and tilting soon after vibration is initiated, i.e. the weak oscillation. As the value of $\Gamma$ increases beyond 2, the segregations occur and exhibit different phases depending on the oscillation frequency. At relatively lower frequency and higher acceleration, the particles fluidize and move up and down violently. Particularly, RBN appears in the higher frequency and acceleration region. In addition, there is a tendency that the higher the oscillation frequency, the longer the segregation process takes place at the fixed acceleration.

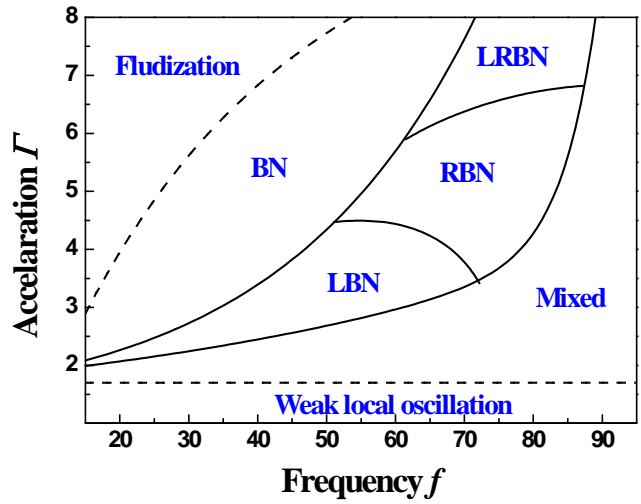

Fig.2. Segregation patterns of the particles for different controlling parameters of the shaker. The binary mixture used consists of 0.24 mm Copper and 0.16 mm glass beads with the same layer thickness. Lines are drawn to help guiding the eye.

We performed many experiments at different frequencies and accelerations to test the behaviors of other particle combinations, and found that the segregation phenomena are rich at higher frequencies and accelerations. Fig. 3 presents the typical phase diagram of segregation at $f = 90\,\text{Hz}$ and $\Gamma = 7$ for various kinds of size and density ratios in binary systems. Each symbol indicates to one experimental data:

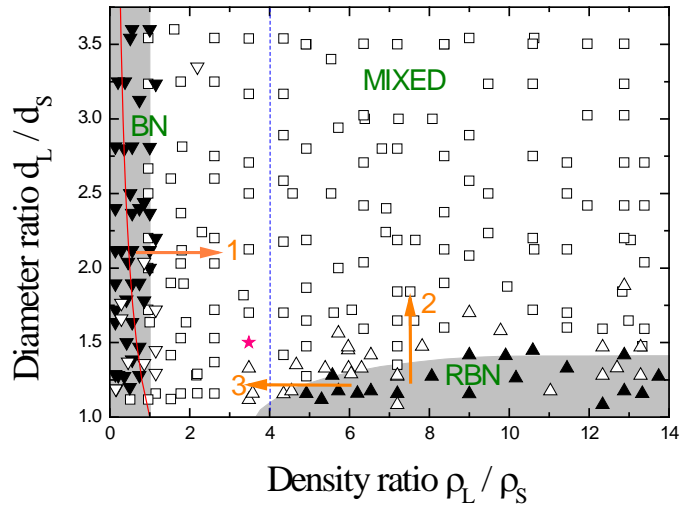

Fig. 3 Segregation patterns of the binary mixtures consisting same-layer thickness: BN (▼), LBN (∇), RBN (▲), LRBN (Δ) and the mixed states (□). The solid line is given by equation (1). Arrow 1, 2, and 3 are respectively corresponding to pattern transition from BNS to Mixed, RBN to Mixed states. The dotted blue line is drawn for comparison with ref. [21] where the biggest density ratio is 4 (the left region of dotted line). The pink star represents the experiment of fig. 2.

close and open down triangles represent BN and LBN states, close and open up triangles stand for RBN and LRBN states, and square denotes to Mixed state, respectively. The gray areas help to guide the eye to approximate regions of BN and RBN. Obviously, BN state is dominant at the region: $\rho_L/\rho_S < 1$, $d_L/d_S > 1.2$ while RBN exists at the region: $\rho_L/\rho_S > 5$, $d_L/d_S < 1.5$. It is noteworthy that in the left region of blue dotted line where $\rho_L/\rho_S < 4$ BN exists, this result is in agreement with the study [21] in which the same layer thickness was not strictly considered. On the other hand, RBN was not observed in this region, but it was reported in ref. 21. This means that the occurrence of BN is independent of layer structure while RBN is sensitive to the layer structure. In fact, RBN phenomenon only occurs as the density ratio is larger than 4 in our system. In addition, we observed a transition from BN (RBN) to Mixed state passes through LBN (LRBN) as shown in Fig. 4 where $a_1a_2a_3$, $b_1b_2b_3$ and $c_1c_2c_3$ are corresponds to arrows 1, 2, and 3 in Fig. 3 respectively. In other words, BN transits to RBN passes through LBN-mixed-LRBN three states, which is partially inconsistence with Hong's theory, i.e., BN transits to RBN passes through a mixed state.

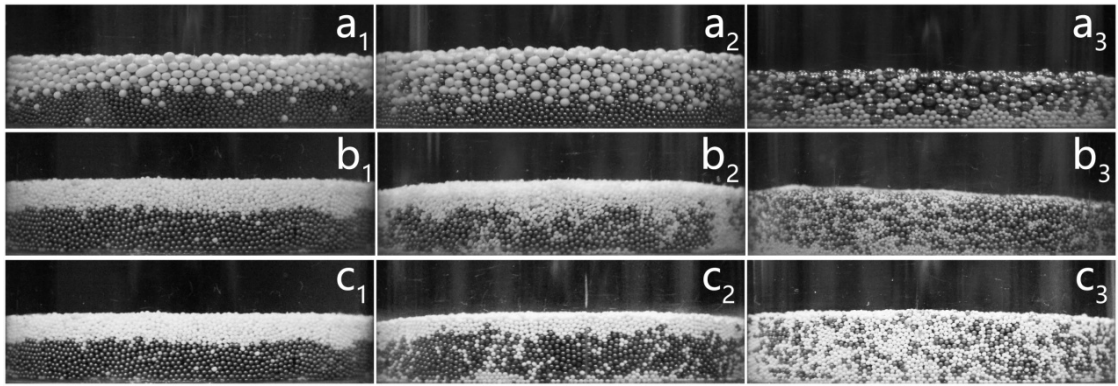

Fig. 4 Transition from BN ($a_1$) to Mixed state ($a_3$) through LBN ($a_2$) (see arrow 1); Transition from RBN ($b_1$) to Mixed state ($b_3$) through LRBN ($b_2$) (see arrow 2); Transition from RBN ($c_1$) to Mixed state ($c_3$) through LRBN ($c_2$) (see arrow 3).

To facilitate comparison with the prediction of equation (1), the borderline for the crossover from BN to RBN is depicted in figure 3 (see the solid line). Obviously, BN and RBN regions are located in the prediction criteria [20]. This indicates that our result is partially consistent with the prediction of Hong et al. However, in our experiment there are wide-range of mixed states are also observed when the density ratio is greater than one. This result is qualitatively similar to the observation of Canul-Chay et al. [22], where BN states existed at low densities and mixed states occurred when the densities of large particles were greater than the small ones, which means BN is the size effect dominating. Obviously, the more the size ratio, the voids become larger, so that the small spheres can easily percolate downward and fill voids below. This means that the void filling mechanism becomes stronger with increasing diameter ratio dL/dS [27]. In contrast with this, in the RBN region where the density ratio is sufficiently higher and diameter ratio is relatively smaller, large heavy particles move down to the bottom and form the lower-layer while small particles rise to the top and form an upper-layer. Hence, RBN can reasonably attribute to the condensation driven [20]. As for the mixed state regime, larger particles are overwhelming to small ones in both sizes. Correspondingly, an increase of diameter ratio results in large volume ratio $V_L/V_S$ , and consequently results in larger voids in the sample, i.e., the void-filling mechanism becomes less effective [27].

## 4. CONCLUSION

The segregation patterns of binary mixtures consisting of same layer-thickness were systematically studied under the vertical vibration. Five kinds of phenomena were observed in the space of size-density ratios: Mixed, BN, RBN, LBN, and LRBN. Furthermore, BN state dominant at lower density ratio and is independent of layer structure, RBN regime exists in the region of higher density ratio and depends on the layer structure, and the mixed pattern occupies most of the area. In addition, LBN seems like to be the transition state from BN to Mixed state, whereas LRBN is the transition from RBN to Mixed state. Our result may be helpful for the establishment of theory about the segregation and mixing of granular mixtures.

ACKNOWLEDGEMENTS

The work was supported by the Chinese National Science Foundation, Project Nos. 10975014, 11274355 and 11475018.